\documentstyle[graphicx,prb,aps]{revtex}

\begin{document}
\date{\today}
\title{Modification of the standard model for the lanthanides}

\author{U.\ Lundin, I.\ Sandalov, O.\ Eriksson, and B.\ Johansson}
\address{Condensed Matter Theory group, Uppsala University, Box 530,
SE-751 21 Uppsala, Sweden}
\maketitle

\begin{abstract}
We show that incorporation of strong electron correlations
into the Kohn-Sham scheme of band structure calculations
leads to a modification of the standard model of the lanthanides and that 
this procedure 
removes the existing discrepancy between theory and experiment
concerning the ground state properties.
Within the picture suggested,
part of the upper Hubbard $f$-band is occupied due to 
conduction band-$f$-mixing interaction (that is renormalized due to 
correlations) and this contributes to the cohesive
energy of the crystal. The lower Hubbard band has zero width and
describes fermionic excitations in the shell of localized $f$-s.
Fully self-consistent calculations (with respect to both charge density and
many-electron population numbers of the $f$-shell)
of the equilibrium volume $V_0$
and the bulk modulus of selected lanthanides
have been performed and a good agreement is obtained. 

\end{abstract}
\pacs{71.15.-m,71.20.Eh,71.27.+a,71.28.+d}

{\bf Published as: Sol.\ State Commun. {\bf 115}, 7-12 (2000).}\\
For strongly correlated systems, state-of-the-art band structure
methods fail to give an accurate description.
The reason for the failure is well-known: the strong intra-shell 
Hubbard repulsion, $U$, 
is underestimated in the calculations.  
If the $f$-$f$ interaction is sufficiently strong, the $f$-shell forms a
localized multiplet (Russel-Saunders coupling) and can 
accurately be treated as atomic like. This
is the basic idea behind the so-called standard model of the
lanthanides and many of the physical properties seem consistent
with it, for example the Curie-Weiss behavior of
the susceptibility~\cite{standard}, the gross features of the equilibrium 
volumes~\cite{volume}, and the structural properties~\cite{struc}.
When using 
{\em ab initio} calculations employing the local density approximation (LDA)
and
treating the $f$-electrons as core states,
a too large equilibrium volume, $V_0$, is obtained (for all other elements,
LDA {\it underestimates} the equilibrium volume).
On the other hand, when the 4$f$'s
are treated as itinerant 
electrons, $V_0$ is much too low compared to experiment and the observed 
localized moment (Curie-Weiss law) is absent. 
Calculations by Delin {\it et al.}~\cite{delin} show that treating the
4$f$-electrons as core states is a very good approximation for the late
lanthanides, but for the lighter lanthanides the
disagreement between theory and experiment is gradually increasing when 
one goes from heavy to the lighter lanthanides. 
Thus, even though the standard model is essentially correct, it has to be 
modified slightly to better describe the cohesive properties of 
the lanthanides.
Such a modification requires a method combining an {\em ab initio} 
band structure calculations (AIBSC) and a many-body approach. A variety of such 
methods are developed.
The ideas of orbital polarization\cite{olle}, LDA+U
method\cite{aza}, Hartree-Fock type of approximations, 
method of phase shift\cite{fulde} and more
advanced methods, like three-body Faddeev equations\cite{fulde,3bFE}, have
in some cases 
been successfully applied to systems with strong electron correlations (SEC).  
New approaches: on the basis of the Gutzwiller wave 
function\cite{zein},
dynamical mean field\cite{kotliar} and the LDA++-approach\cite{lichtenstein}
have also been used for the description of different SEC systems. 
In many cases the AIBSC-based density of electron states 
is used as input to a model calculation using a non-self-consistent 
fixed lattice parameter and Hubbard $U$. 
However, since a large value of Hubbard repulsion restricts the 
available electron phase space, wave functions and, correspondingly, the 
matrix elements of mixing and hopping, changes.
Recently an attempt to describe SEC by means of slave-boson method\cite{italy} 
(i.e. Hubbard $U=\infty $) build into the scheme of linear
muffin-tin orbital method (LMTO) was made. The author 
found that the
results are unsatisfactory\cite{italy} and  had to use a scheme 
different
from the standard slave-boson self-consistency schemes. 
However, in all these attempts the influence of SEC on
the cohesive properties has not been questioned yet.
Our approach to improve on the standard model is based on the following 
physical picture: the Hubbard repulsion is sufficiently large to 
form large energy gaps between the $(n-1)$-, $n$- and $(n+1)$-electron 
configurations not only in the heavy but also in the light lanthanides . 
Near the atomic limit {\em all} $f$-electrons are in 
the {\em strong} coupling regime 
and hopping and mixing interactions 
lead to the formation of a lower Hubbard sub-bands 
(LHB, $|n-1\rangle \rightarrow |n\rangle$ transitions) and a upper sub-band
(UHB, $|n\rangle \rightarrow |n+1\rangle$ transitions).
In our case, however, the lower transitions are substantially 
below the bottom of the conduction bands (Ce is an exception) and, hence, 
they do not experience any mixing interaction. 
Hubbard $U$ increases with atomic number, due to the more localized
nature of the 4$f$'s, thus the LHB goes down in energy while the UHB
goes up and therefore it experiences smaller mixing and is populated 
with fewer states. 
These many-electron local excitations correspond to 
localized core electrons within the one-electron picture. 
The upper transitions are 
in the energy region where the delocalized electrons do exist. Therefore, 
the overlap and mixing interaction may delocalize these transitions, forming 
an UHB, filled only by a small amount, $\eta$, of electrons. 
This picture corresponds to the multi-orbital Hubbard-Anderson model. 
 The total occupation number can be written as
$n_f = n + \eta$, where $n$ is an integer, which is determined by the nuclear 
charge valence, and $0<\eta < 1$. 
Thus, most of the $f$-occupation is in the LHB. This is consistent
with the observed 
Curie-Weiss behavior of the high-temperature (above T$_c$) susceptibility. 
In the lanthanides the width of the appropriate $f$-sub band 
is much smaller than the Hubbard band gap. This allows one
to formulate an effective Schr\"odinger (or Dirac) equation (SE) for two 
different energy regions. 
As is known from Hubbard's models\cite{hubb},  the single-electron 
spectral weight at $U=0$ is split by the Hubbard repulsion $U$ to a set of
weights, corresponding to an electron transition
from one ion state to another. These weights renormalize the matrix
elements of the hopping and mixing interaction, which in turn,
cause a narrowing of the bands due to correlations.
The theory is developed for a non-orthogonal 
basis set, as for instance 
used in the linear muffin-tin orbital method in the atomic
sphere approximation (LMTO-ASA)\cite{lmto}.
Here we apply the formalism only in the lowest approximation which 
requires less complicated changes for methods
used in self consistent {\em ab initio} calculation.
Compared to conventional considerations of the periodical Anderson
model (PAM) which do not include structure-dependent vertex corrections 
we solve
the Schr\"odinger equation for the electrons in fields generated by 
the nuclei and electrons in every iteration,
with the charge density obtained from a system of equations
for Greens functions (GF) and spectral weights, thus renormalizing 
self-consistently the
matrix elements of the PAM. On the other hand compared to 
the normal LDA approach we solve self-consistently additional
 system of equations for many-electron spectral weights and obtain a
bandwidth which is reduced by correlations. 

Technically our method, briefly described, 
consists of the following steps: \\
 1. The total many-electron Hamiltonian is written in terms of 
an LMTO basis set.\\
2. The $f$-electron operators are transformed into a Hubbard representation.\\
3. Analytical calculations are performed for the electron Greens
functions for the corresponding PAM in the LMTO
representation in the Hubbard-I 
approximation\cite{hubb} (strong-coupling regime).\\
4. Renormalizing factors to the $f$-$(spd)$-blocks of the
LMTO Hamiltonian and overlap matrices are introduced which are derived 
from a comparison of the 
frequency and overlap matrixes. They arise due to strong correlations 
in the system for Greens functions. \\
5. Self-consistent, {\em ab initio} calculations of 
the system with SEC are performed. \\
In practice, the main effect of
these steps consists of the separation\cite{kuzmin} of the 
$f$-electron system into two subsystems
and the correlation driven narrowing of the $f$-bands in the conventional LMTO 
Hamiltonian.
This, as will be demonstrated below, leads to an improved description of
the bonding of the lanthanide metals.
In order to describe our theory in more detail, the secondary 
quantized full many-electron 
Hamiltonian, ${\mathcal H}$, is written 
in terms of an LMTO-basis set. 
An unperturbed Hamiltonian and a perturbation is formulated as
$ {\mathcal H} =  (T+V_{ne}+V_{LDA}) + (H_{ee}-V_{LDA})
={\mathcal H}_0 + {\mathcal H}'. $
The main effect of the $f$-part single site correction from Coulomb
interaction, 
\begin{equation}
{\mathcal H}'_U =  \frac{1}{2} \sum U_{m_1 m_2 m_3 m_4}
 f^{\dagger}_{m_1 \sigma} f^{\dagger}_{m_2 \sigma'}
f_{m_3 \sigma'} f_{m_4 \sigma}
- \sum v_{l=3,\sigma}^{(LDA)} \sum
f^{\dagger}_{m_l \sigma} f_{m_l \sigma},
\label{lmtoham}
\end{equation}
is to localize part of the $f$-spectra,
similar to a so called LDA+U solution.
Let us denote all non-$f$-operators in the LMTO representation as
$c_{{\bf k}L}$. 
Then the LMTO part of the Hamiltonian, $\sum H_{LL'}({\bf k})
a^{\dagger}_{{\bf k}L}a_{{\bf k}L'}$, can be written as a sum of
$s,p,d,$ and $f$-electrons, and their mixing interaction,
\begin{eqnarray}
& & {\mathcal H}_0 = 
\sum _{{\bf n} m_l \sigma} \epsilon_f^0 f^{\dagger}_{{\bf n}m_l \sigma}
f_{{\bf n}m_l \sigma}+
\sum _{{\bf k}L,L'} H_{L,L'}({\bf k})
c^{\dagger}_{{\bf k}L}c_{{\bf k}L'} + \nonumber \\
& & \sum _{{\bf nm},m_l \sigma m'_l}
t^{m_l  m'_l}_{{\bf nm}} 
f^{\dagger}_{{\bf n} m_l \sigma }f_{{\bf m} m'_l \sigma } +  
\sum _{{\bf k,n},L,m_l\sigma} [H_{L,m_l\sigma}({\bf k}) e^{i{\bf kR_n}}
c^{\dagger}_{{\bf k}L\sigma}f_{{\bf n}m_l\sigma} + h.c.],
\label{starthamiltonian}
\end{eqnarray}
where 
$ t^{m_l  m'_l}_{{\bf nm}} = \sum _{\bf k} e^{i{\bf k(R_n - R_m)}} 
[H_{m_l \sigma, m'_l \sigma}({\bf k}) - \delta_{m_l,m'_l}\epsilon_f^0]
$
is the hopping matrix element between $f$-orbitals and
$\epsilon_f^0 = \sum_{{\bf k}} H_{m_l\sigma,m_l\sigma}({\bf k})$.
Since the LMTO-ASA basis is non-orthogonal
with an overlap matrix
$\langle\chi _{{\bf n}L} | \chi _{{\bf m}L'}\rangle = 
{\mathcal O}_{{\bf n}L,{\bf m}L'}$,
the anticommutator is $\{a_{{\bf n}L}, a_{{\bf m}L'}^{\dagger} \} =
({\mathcal O}^{-1})_{{\bf n}L,{\bf m}L'}$. Therefore, the equation for the
bare fermion Greens function, 
$G_{{\bf n}L,{\bf m}L'}(\omega ) \equiv 
- \langle T[ a_{{\bf n}L}(t) 
a_{{\bf m}L'}^{\dagger}(0)]\rangle_{\omega},$
for this Hamiltonian is
\begin{equation}
[\omega {\mathcal O}_{{\bf n}L,{\bf m_1}L_1} - 
 H_{{\bf n}L,{\bf m_1}L_1}]
G_{{\bf m_1},L_1,{\bf m}L'}(\omega ) = \delta _{{\bf n},{\bf m}}
\delta _{L,L'}. \nonumber
\end{equation}
Thus, we find (by construction) in the square brackets the 
conventional secular problem for
the LMTO-ASA method (or any LDA based method).
Introducing a full set of states in an orbital representation
$|\Gamma_0)=|0\rangle, |\Gamma_1)=|\gamma\rangle, |\Gamma_2)=|\gamma,
\gamma'\rangle,\cdots$
($\gamma$ denoting a single particle $f$-state, $m_l\sigma$)
allows to express
any single-site operator $\hat{{\mathcal A}}$ in terms of $X$-operators,
$\hat{{\mathcal A}} = \sum_{\Gamma,\Gamma'}
\langle\Gamma|\hat{{\mathcal A}}|\Gamma'\rangle X^{\Gamma,\Gamma'}$, 
$X^{pq} = |p\rangle\langle q|$ and 
where $\langle X^{\Gamma, \Gamma}\rangle
\equiv N_{\Gamma}$ is the occupation number for the state $\Gamma$.
Now we can express the zero $f$-Greens functions, for the LMTO problem, 
in terms of $X$-Greens functions as
$G^{(0)}_{\gamma}(\omega) = \sum_{a} |(f_{\gamma})_{a} |^2
P_{a} D_{a}(\omega),$
where
$D_{a}(\omega) = \frac{1}{ \omega - \Delta_{a} },$
and an $X$-operator GF ${\mathcal G}^{a{\bar a}}=P^a D_a$. 
Here $a=(\Gamma_n,\Gamma_{n+1})$ (${\bar a}=(\Gamma_{n+1},\Gamma_n)$) 
are possible Hubbard transitions,
$P_{\Gamma_n,\Gamma_{n+1}}=N_{\Gamma_n} + N_{\Gamma_{n+1}}$,
$\Delta_{\Gamma_{n+1},\Gamma_{n}} = E_{\Gamma_{n+1}}-E_{\Gamma_{n}} -
\mu $,
 $E_{\Gamma_{n}} \simeq n\epsilon_f^0 +U n(n-1)/2$ and $\mu$ is the chemical
potential.

The $f$-electron operators can be expanded in the 
Hamiltonian ${\mathcal H}$ into $X$-operators for
the set of many-electron states described above. 
Then we arrive at an Anderson like model 
\begin{eqnarray}
{\mathcal H} = \sum _{{\bf n}\Gamma}
E_{\Gamma } X^{\Gamma \Gamma }_{{\bf n}} +
\sum _{{\bf k},L,L'} H_{L,L'}({\bf k})
c^{\dag}_{{\bf k}L}c_{{\bf k}L'} +  \\ \nonumber
\sum _{{\bf k,n},\mu,a}
[H_{L,\mu}({\bf k}) (f_{\mu})_{a} e^{i{\bf kR_n}}
c^{\dag}_{{\bf k}L}X_{{\bf n}}^{a} + h.c.]. + 
\sum_{{\bf n,m},a,b} t_{\bf nm}^{ab} X_{{\bf n}}^{a} X_{{\bf m}}^{b}.
\nonumber
\end{eqnarray}
Here $t_{\bf nm}^{ab}=\sum_{\mu,\nu} (f^{\dag}_{{\bf n}\mu})^{a} 
(f^{\dag}_{{\bf m}\nu})^{b} t_{\bf nm}^{\mu\nu}$.
For the derivation of the equations for the Greens functions, 
${\mathcal G}^{(cX)} \equiv
-\langle T[ c_{{\bf n}L}(t) X^{M\Gamma}_{{\bf n}}(0)]\rangle$ and
${\mathcal G}^{(XX)} \equiv
-\langle T[ X^{\Gamma M}_{{\bf n}}(t) 
X^{M\Gamma}_{{\bf n}}(0)]\rangle$,
(which will be compared to the equation for the regular 
Kohn-Sham (LMTO) problem) 
we need the anticommutators $\{c,X\}$, we find  
$\{ c_{{\bf n}L},X^{\Gamma'\Gamma}_{{\bf n'}}\} =  \\ \sum_{{\bf n_1} \mu
\Gamma_1 \Gamma_2} {\mathcal O}^{-1}_{{\bf n}L,{\bf n_1} \mu}
(f_{\mu})_{\Gamma_1,\Gamma_2}
\{ X_{{\bf n_1}}^{\Gamma_1,\Gamma_2}, X_{{\bf n'}}^{\Gamma',\Gamma} \}.$
Physically, this implies that the $f$-part 
of the conduction-band wavefunction, coming 
from other sites, experiences strong correlations on the reference ion.
The Hubbard-I as well as mean field approximation correspond to
$\{ X_{{\bf n}}^{\Gamma_1,\Gamma_2}, X_{{\bf n}}^{\Gamma',\Gamma} \}
\simeq
 P_{\Gamma',\Gamma} \delta_{\Gamma',\Gamma_2}\delta_{\Gamma,\Gamma_1}$.
Introducing renormalized "Fermi" operators 
$\tilde{f}^{\Gamma',\Gamma} = X^{\Gamma',\Gamma}/\sqrt{P_{\Gamma',\Gamma}}.$
We see that the matrix form of the
GF, ${\mathcal G}^{(ij)}$, with $i,j = c,X$, coincides with the GF of the
Kohn-Sham equation, $G$,
with the only difference being $A_{L,\mu} \rightarrow \tilde{A}_{L,\mu}=
\sum_{\Gamma',\Gamma} A_{L,\mu} (f_{\mu})_{\Gamma',\Gamma} \sqrt{
P_{\Gamma',\Gamma}}$, where $A = H$ or ${\mathcal O}$. 
Thus the most essential, technical part of
our theory leads to an effective secular equation which, when
diagonalized gives eigenvalues with many-body corrections.

As an example let us consider praseodymium metal, which has two localized
$f$-electrons. 
Therefore, $|\epsilon_f^0|$ has a value which gives an energy minimum for $n=2$.
For briefness we will use the following notation for the $f$-orbitals:
$\gamma = (m_l=3,\downarrow) \equiv 1, (m_l=2,\downarrow)
\equiv 2$ and $\gamma=\nu$ for $ \gamma \neq 1,2$.
$\Gamma_{(12)}\equiv (12)$, $\Gamma_{(12\nu)}\equiv (12\nu)$.
Let us now consider a simple polarized solution, where
the $f$-orbitals with $\gamma = 1,2$
are fully occupied, while the rest of the $f$-electrons occupy
the $f^3$-states, $|\Gamma_3\rangle = |1,2,\nu\rangle$. When mixing is absent,
$n_f = 2$ and $N^0_{(12)}=1$ whereas all other population
numbers, $N^0_{\Gamma}$, are zero.
Then ${\mathcal G}^0_{\gamma =1}= D_{2(12)}, 
{\mathcal G}^0_{\gamma =2}= D_{1(12)} $ and
${\mathcal G}^0_{\gamma =\nu}= D_{(12)(12\nu)}.$
We denote
$\Delta_{(12)2} = \Delta_{(12)1} \equiv \Delta_1,$ and
$\Delta_{(12\nu)(12)} \equiv \Delta_2$.
In the limiting case, $\Delta_2,\Delta_3 \rightarrow \infty$, no $f$'s are 
present in the conduction band, and therefore this corresponds exactly to 
the limit of the standard model. 
The $f_{\nu}$-bands are slightly
above the
Fermi energy, and  mixing interaction transfers $f$-character into the
$s,p,d$-electron states.
In a general case it is, of course, impossible to
write equations for $N_{\Gamma}$ in terms of only orbital GF's,
${\mathcal G}_{\gamma}$, because equations for higher correlation
functions
are needed. However, if for simplicity
we assume that these 12 $\nu$-bands are occupied
symmetrically, {\em i.e.}, $N_{(12\nu)} = (1/12)\sum_{\nu'}N_{(12\nu')}$,
we immediately find
\begin{equation}
\left\{
\begin{array}{l}
n_f = 2\cdot N_{(12)} + 3 \cdot \sum_{\nu} N_{(12\nu)} = 2 + \eta,  \\
N_{(12)} + \sum_{\nu} N_{(12\nu)} = 1, \nonumber
\end{array}
\right.
\end{equation}
i.e., $N_{(12\nu)}=\eta/12$ and $N_{(12)} = 1- \eta$ (for the other rare-earth
elements a slight modification of these expressions are needed).
Thus, renormalization of the population numbers leads to the form of
$f$-locators
(single-site, ${\bf k}$-independent part of the GF)
\begin{equation}
{\mathcal G}_1 ={\mathcal G}_2 = \frac{1-\eta}{\omega - \Delta_1} +
\frac{\eta}{\omega - \Delta_2}
\mbox{\hspace{0.2cm} and \hspace{0.2cm}}
{\mathcal G}_{\nu} = \frac{1-11\eta/12}{\omega - \Delta_2} +
\frac{11\eta/12}{\omega- \Delta_3},
\label{gnu:eqn}
\end{equation}
where $\Delta_3 \equiv \Delta_{(12\nu\nu')(12\nu')}$. The center of
the non-renormalized
LDA-$f$-band is in $\omega = \Delta_2$ and ${\mathcal G}_1^{LDA} =
{\mathcal G}_2^{LDA} = {\mathcal G}_{\nu}^{LDA} = 1/(\omega- \Delta_2).$
Let us ignore for the moment the non-orthogonality.
For a diagonal effective hopping
$t_{\gamma}({\bf k},\omega)$, of any origin, we have
$G_{\gamma} = [(G_{0,\gamma}^{(at)})^{-1} - t_{\gamma}]^{-1}$.
Therefore, the expressions for the local, $f$-part, of the self-energies are
$\Sigma _{\gamma} = ( G_{\gamma}^{LDA})^{-1}
 - U\sum_{\gamma'}(1 - \delta_{\gamma, \gamma'})n_{\gamma'}
 - (G_{0,\gamma}^{(at)})^{-1} $.
Since all poles in ${\mathcal G}_{\gamma}$ are well separated, 
one can formulate two
effective
Schr\"odinger equations near $\omega \sim \Delta_1$ and
$\omega \sim \Delta_2$. Renormalization factors
$(1 - (\frac{\partial \Sigma}{\partial \omega})_{\omega = \Delta_i} )$
are,
of course, given by spectral weights (numerators) of the GF's in these
poles. Since
$\eta$ is small the bandwidths of the upper Hubbard sub bands for
$\gamma = 1,2$ are also small and, being above the Fermi level, they are
empty. Within the scenario considered 
the lower sub bands have weight slightly less than one, $(1-\eta)$.
For the bands $\nu$, $\Delta_3$
is far above the Fermi energy. Thus the upper sub bands are empty and
the spectral weight
in the pole $\omega = \Delta_2$ is $(1 - \frac{11\eta}{12})$.
Therefore, the effective
Schr\"odinger equation for bands $1$ and $2$ at 
$\omega \sim \Delta_1 = \Delta_2 - U$ gives
$\omega - (\Delta_2 - U) - (1 - \eta) t_1({\bf k},\omega) = 0$,
whereas for the bands $\nu$ at $\omega \sim \Delta_2$ we find $\omega -
\Delta_2 -(1 - 11\eta/12) t_{\nu}({\bf k},\omega) = 0$.\\
Since in the case of LDA calculations with delocalized $f$-electrons 
the center of the $f$-band is $\sim \Delta_2$ 
we conclude that
the potential for the 1st and 2nd orbitals is shifted down by $U$
(similar to the LDA+U - result,\cite{lda+u} and self interaction
correction (SIC)-theory \cite{sicpr} 
but {\em without} factors $n_{\mu}$). The next conclusion is that 
the described factors should be to introduced 
into the LMTO overlap and Hamiltonian matrices. Thus, we calculate the 
charge density and a self consistent potential,
corresponding to the given $\eta$, 
where 
$\eta$ comes from the equation of self-consistency
\begin{equation}
\label{eq:eta}
\eta=-\frac{1}{\pi}\sum_{\omega,\nu,{\bf k} }  f(\omega) Im 
{\mathcal G}^{(12,12\nu),(12\nu,12)}_\nu(\omega+i\delta,{\bf k}),
\end{equation}
since the parameter $P_{\Gamma',\Gamma}=1-\frac{11}{12}\eta$ renormalizes 
${\mathcal H}$ and ${\mathcal O}$.
Here ${\mathcal G}^{(12,12\nu),(12\nu,12)}_\nu$ 
is obtained from the transformed secular matrix for the normal LMTO-GF's,
$(\tilde{{\mathcal O}}E - \tilde{{\mathcal H}})^{-1}$, via $\tilde{f}$. 
For Pr we find the following picture. Two of the
f-orbitals are located at an energy -$U$ lower than the remaining $f$-states 
and below the bottom of the conduction bands. Therefore, 
we treat those $f$-states as core states removing
the somewhat arbitrary choice of a value for the Hubbard $U$.
Compared to LDA+U and SIC methods, in our approach 
the spd-f hybridization is reduced
by the renormalizing factor $\sqrt{P}$. 
The theory is generalized for the other lanthanides straightforwardly.

Before giving numerical examples of cohesive properties (e.g.
equilibrium volume) obtained from our method
we describe how the total energy was calculated.
As is well-known, the equation for the exchange-correlation potential
$v_{xc}$ 
derived by Sham\cite{sham} connects it with the exact self-energy
of the electron system. Making use of a 
strong-coupling (SC) perturbation theory (PT) developed by us 
we have performed the analysis of the contributions to the self-energy 
and found that a there exist a one-to-one correspondence between the graphs for 
the self-energy in the 
standard weak-coupling (WC) PT (in e.g. the random phase approximation) and the
sequence of graphs in SCPT. 
Using the facts that: 
a) Sham's equation\cite{sham} should produce a $v_{xc}$ corresponding
to the particular SE choice;
b) established correspondence between the WC and SC PT graphs;
c) the statement of Kotani\cite{kotani}, that the static random phase 
approximation reproduces well the DFT-LDA calculation with the standard
choice of $v_{xc}$, we come to the conclusion that, within 
the approximation chosen, the original LDA functional form for $v_{xc}$ can be 
used. However, the charge density is calculated from renormalized
fermions which differ from the standard fermions by the renormalizing factors
$\sqrt{P}$. Hence the charge density,$\rho_P(r)$, also depends on 
$\sqrt{P}$. The mixing and hopping  in the effective Kohn-Sham 
equation should be renormalized by the same $P$ as described above  
and a part of the renormalized $f$-electrons should be described as localized 
(lower transitions ) and 
the rest (coming from the upper transitions ) 
as delocalized. The separation is caused by self-interaction correction 
automatically generated in PT. Finally, 
the equation of self-consistency for parameter $P$ (Eqn~\ref{eq:eta}) 
should be added.

We have hence 
calculated the total energy using the regular LDA total
energy expression, using the electron density $\rho_P(r)$, i.e.
$E_{LDA}(\rho_P(r)$, where the parameter $P$ which comes from electron
correlations, is found self-consistently from additional equation and 
regulates distribution of spectral weight between lower and upper parts
of energy.
\begin{figure}
\includegraphics[width=0.9\textwidth]{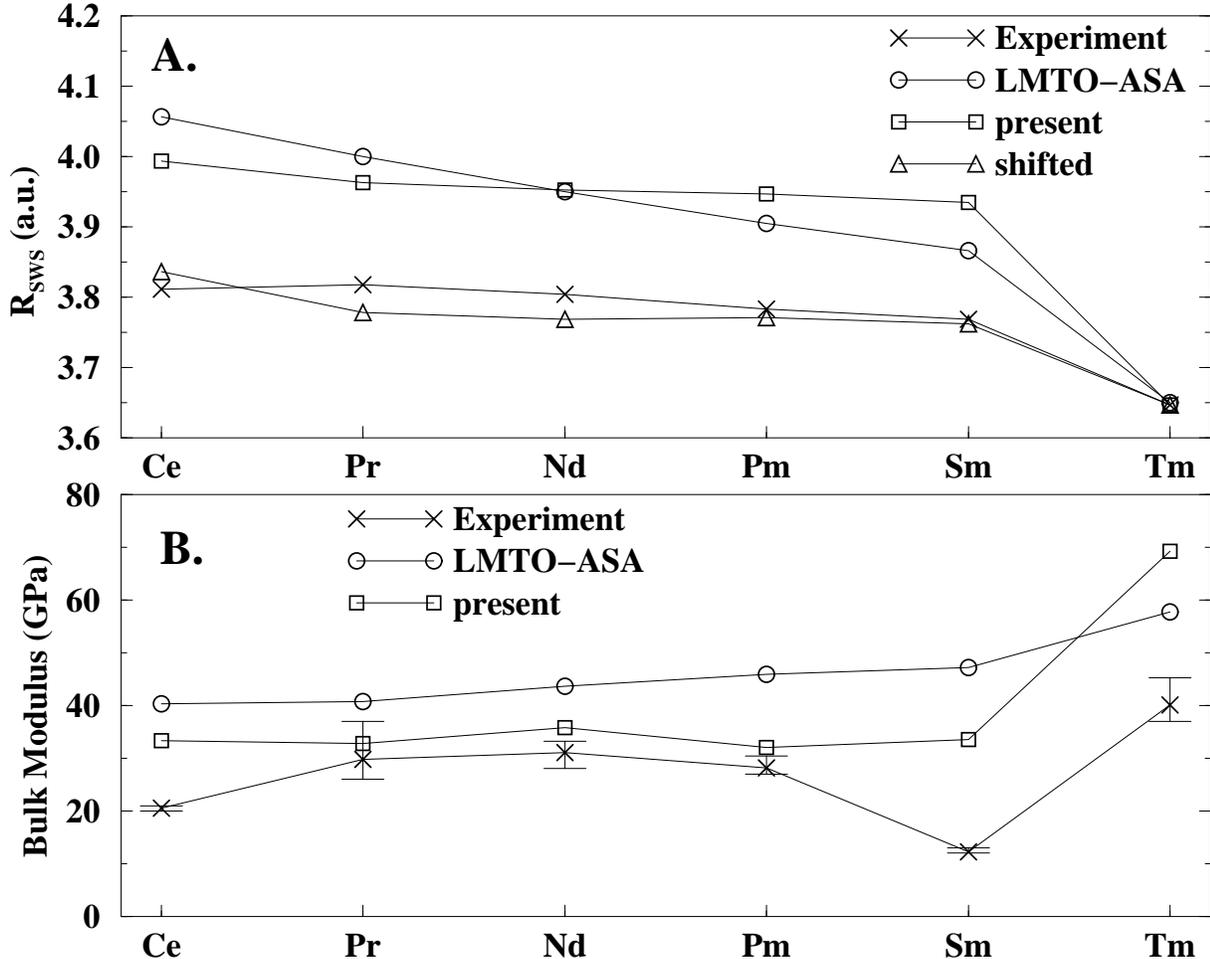}
\caption{Fig. A shows the equilibrium volumes, in 
experiment, LMTO-ASA (using the standard model), 
corrected LMTO-ASA and shifted theory, for some selected lanthanides.
The basis-set shifts used are (in a.u.) 
for Ce $\Delta R_{sws}$=-0.157, -0.185 for Pr, -0.184 for Nd, 
-0.176 for Pm, -0.172 for Sm and -0.003 for Tm.
Fig. B shows the bulk modulus for the same elements.
The slope of the lanthanide correction is reduced due to the contribution
to the cohesive energy from the delocalized $f$'s.}
\label{fig:vol}
\end{figure}
In Fig.~\ref{fig:vol}A we show the equilibrium volumes of some selected 
lanthanides, from experiment, uncorrected Kohn-Sham 
calculations (LMTO-ASA) and from our 
corrected theory. Note that the many-body corrections are presented 
together with shifted data  due to an incomplete basis set used 
in the LMTO-ASA calculations. This shift was calculated by comparing (LDA) 
volumes obtained from a full-potential-(FP)-LMTO, multi-basis set 
calculation with 
(LDA) volumes from LMTO-ASA calculations, 
$\Delta R_{sws}=R_0^{FP}-R_0^{LMTO-ASA}$. 
An overall good agreement is obtained, and for the heavier lanthanides the 
renormalization correction becomes vanishingly small (as seen for Tm).
The only parameter of our model, the position of the UHB with respect to the 
Fermi level, was taken from bremsstrahlung isochromat spectra (BIS)~\cite{baer}.
In principle this value can be found from a super cell calculation changing the 
$f$-occupancy at one site.
For cerium, $\Delta_{n,n+1}$ is $\simeq$3.5 eV above the 
Fermi level, and 
shifting this to 3.5$\pm$ 1 eV, only affected the equilibrium volume with 
$\pm$0.5\%.
For samarium the position of the UHB is $\simeq$0.6 eV, and changes in 
the position with $\pm$0.1 eV, changed the volume with $\pm$0.8\%. 
For the other elements changes were of the same order. 
Therefore, we conclude that the results are rather insensitive
with respect to the position of the UHB.
Fig.~\ref{fig:vol}B shows the bulk modulus. Note that this value is quite 
sensitive to the fitting procedure (pressure-volume fit) and to the
structure, thus, it should be 
considered only as an approximate value. 

The physical picture obtained 
is consistent with previous theoretical work
on Fermi surfaces~\cite{norman} where a better agreement 
between theory and experiment was obtained when including, ad-hoc, the 
4f states as valence electrons
even though part of these 4f states were treated as occupied core states.
It is also consistent with results obtained from self interaction corrected
density functionals on Pr metal\cite{sicpr}. Unlike the method of orbital 
polarization the present theory is a true {\em many-body} correction, 
and thus presents a different physical approach. 
The decisive parameter in our case (unlike to the LDA+U method) 
is not the absolute value of the Hubbard $U$ but the  position 
of the transition $E(\Gamma_{n+1})-E(\Gamma_{n})$ with respect to 
Fermi energy. We have taken this parameter from experimental data~\cite{baer} 
and this led us to an improvement in the calculated cohesive properties. 
First loop corrections, affecting the position of the levels, are small in 
our case~\cite{lundin}.
Fluctuations coming from charge excitations are expected to 
alter the situation at non-zero temperature, however, this 
remains an open question.

We have shown that many-body corrections to the standard model 
improves the description of the cohesive properties of the lanthanides, and
explains the observed discrepancy for the equilibrium volumes
for the light lanthanides. 
The only input to the present theory is the position of the upper
Hubbard transition, $\Delta_{\Gamma+1,\Gamma}$, which was taken from 
bremsstrahlung isochromat spectra (BIS) 
data. We found that a moderate change in energy of this transition gives
only a minor change in the equilibrium volume. The calculated 
volumes are in good agreement with the experimental ones,
which means that the excited states contributes to the ground state 
properties.
The presented theory is designed to deal with strongly 
correlated electron systems and is integrated in an electronic structure 
method, and made fully self-consistent. It describes an additional 
contribution to the chemical bonding from a fraction $f$-electrons 
hybridized with the conduction electrons. 
Compared to for instance the LDA+U, orbital polarization or SIC methods
the present theory involves a 
decrease of mixing due to strong correlations, causing 
the $f$-bandwidth to be reduced.

We are thankful for financial support from
the Swedish Natural Science Research Council.
The critical reading of our manuscript by Prof.\ Helmut Eschrig is
greatly appreciated.


\end{document}